\documentclass{PoS}
\def\rW{\ensuremath{\mathrm W}}
\def\rWW{\ensuremath{\mathrm W{\mathrm W}}}
\def\lss{\pm\,\pm}
\def\col{Collaboration}
\def\ea{{\it et al.}}

\newcommand{\ZPC}[3]  {Z.\ Phys.\ \textbf{C#1} (#2) #3}

\newcommand{\PRL}[3]  {Phys.\ Rev.\ Lett.\ \textbf{#1} (#2) #3}
\newcommand{\CPC}[3]  {Comput.\ Phys.\ Commun.\ \textbf{#1} (#2) #3}

\PoS{PoS(HEP2005)304}

\title{Bose-Einstein correlations and color reconnection in hadronic W}

\ShortTitle{Bose-Einstein correlations and color reconnection in hadronic W}

\author{\speaker{Raimund Str\"ohmer}\thanks{For the LEP Collaborations.}\\
        LMU Muenchen, Germany\\
        E-mail: \email{Raimund.Stroehmer@Physik.Uni-Muenchen.de}}

\usepackage{epsfig}

\abstract{We report on studies of Bose-Einstein correlations and 
color reconnection in hadronic W pair production at LEP.
 Bose-Einstein correlations between identical particles from
 the decay of different W's are studied by comparing the particle
correlation in hadronic W pair decays to those in events which are
constructed from the combination of the hadronic parts of two events
where one of the Ws decays leptonicaly. The LEP combined result of the
strength of the effect is consistent with zero and can be used to
limit the systematic uncertainty due to Bose-Einstein correlations
on the W mass measurements. 

 Color reconnection is expected to affect the production of
 particles in hadronic decays of W pairs.  Measurements of
 inclusive charged particle multiplicities, and of their angular
 distribution with respect to the four jet axes of the events, are
 used to test models of color reconnection.
 The results are both consistent with models without color reconnection
and models with moderate color reconnection. They can be used to
exclude more extreme models of color reconnection and thereby limit
the uncertainty due to color reconnection on the W mass measurement.
 }

\FullConference{International Europhysics Conference on High Energy Physics\\
		 July 21st - 27th 2005\\
		 Lisboa, Portugal}

\begin{document}

\section{Introduction}
In W pair events both W bosons can decay hadronically. In this case
two strongly interacting color singlet objects with considerable space
time overlap are produced. It is interesting to study whether the
the two W bosons then decay independently or whether they influence each
other. In addition to the principle interest in this question 
possible interactions and correlations between the decay product
of the two W bosons influence the precision measurement of the W mass
at LEP and are treated in the mass measurement as source of 
systematic uncertainty.

\section{Bose-Einstein correlations}
Bose-Einstein correlations (BEC) between identical bosons are a 
well-known phenomenon in high energy physics~\cite{BECinHEP}.
BEC lead to an enhancement of the production of identical bosons close in
phase space.
At LEP BEC have been unambiguously established between the particles 
originating from one  hadronically decaying W, representing 
so-called {\it intra}--W BEC \cite{lep-wbec}.
The possible additional BEC  between the particles originating 
from {\it different} W bosons, i.e. {\it inter}--WW BEC, has
to be separated from the {\it intra}--W BEC.
BEC are usually presented in terms of two-particle densities, 
$\rho_2(Q)$, measured as
\[
\rho_2(Q) =
\frac{1}{N_{\rm events}}\frac{\mathrm{d}N_{\rm pairs}}{\mathrm{d}Q}\:,
\label{rho2}
\]
for the number $N_{\rm pairs}$
of pairs 
of identical bosons with four-momenta $p_1$
and
$p_2$ and
$Q=\sqrt{- (p_1 - p_2)^2}$ 
in the number $N_{\rm events}$ of events under study.  

The observed two-particle densities in the W pair decays is then 
compared with the expectation  in the case of non {\it inter}--WW BEC 
which is constructed from W pairs where only one W decays as:
\[
 2\,\rho_2^{\rW}(Q) + 2\,\rho_{\rm mix}^{\rm
WW}(Q)\,.
\]
The second term is a measure of the two-particle densities between
particles from different W bosons which decay independently. Is is
estimated by artificially mixing the hadronic part of two semileptonic
W pair events. Care has to be taken that existing correlations
between the two W bosons are represented in the mixed events.
Only events which polar angles pointing to the same detector
region (e.g. 
$ \left|\theta_{\rW^+} - \theta_{\rW^-}\right| \le 75\,\, {\rm mrad}$
or $ \left|(\pi-\theta_{\rW^+}) - \theta_{\rW^-}\right| \le 75\,\,\mathrm{mrad}$)
are combined and one event is then rotated so that both event are back to back.
The standard selection for hadronic W pair selection is then applied to the
mixed events. Monte Carlo simulations and distributions not sensitive to BEC are
used to establish that these mixed events represent uncorrelated W pair events.
In order to maximize the sensitivity and to minimize systematic uncertainties different
ways to compare the measured two-particle density with that estimated for events
without {\it inter}--WW BEC have been used.
{ \small
\[
 \Delta \rho(Q) = \rho_2^{\rWW}(Q) - 2\,\rho_2^{\rW}(Q) -
 2\,\rho_{\rm mix}^{\rm WW}(Q)\,,
\quad
\delta_I (Q) = \Delta \rho(Q) /  \rho_{\rm mix}^{\rWW}(Q)\,,
\quad
D(Q) = \frac{\rho_2^{\rWW}(Q)}{ 2\,\rho_2^{\rW}(Q) + 2\,\rho_{\rm 
mix}^{\rm
WW}(Q)}\,
\]
}
By either comparing the result for the data with the expectation from
Monte Carlo without BEC or by comparing the result from same-sign
pairs  with opposite-sign pairs uncertainties due to the mixing procedure
can be reduced.
\[
\Delta \rho'(Q) =\Delta \rho(Q) - =\Delta \rho_{\rm no-BEC\: MC}(Q)\,, 
\quad
D'(Q) = \frac{D(Q)}{D_{\rm no-BEC\: MC}(Q)}
\]
\[
d(Q)= D(\lss)/D(+\,-)\,.
\quad R^*=\left({N_{\pi}^{++,--}}/{N_{\pi}^{+-}}\right)^{\rm data}/
          \left({N_{\pi}^{++,--}}/{N_{\pi}^{+-}}\right)^{MC}_{\rm no-BEC}
\]
\begin{figure}
\begin{center}
\epsfig{file=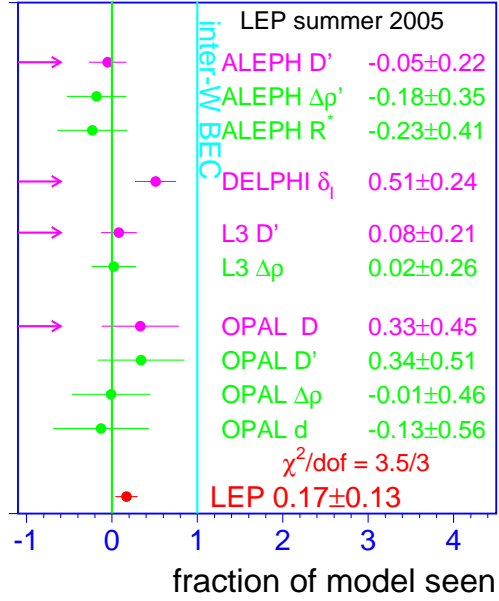,width=8cm} 
\end{center}
\caption{Relative strength of {\it inter}--W BEC compared to the
strength of  {\it intra}--W BEC. The measurements marked with arrows are used
for the combination.
 \label{fig-1}}
\end{figure}
The measured distributions are then compared to Monte Carlo simulations with 
{\it inter}--W BEC correlation. The strength of that correlation relative to the
strength of {\it intra}--W BEC is then varied to determine which scaling factor
best describes the measurements. 
Results for different observables from the four LEP experiments together with a combination
are shown  in Figure~\ref{fig-1}\cite{lep-result}. No significant evidence for {\it inter}--W BEC correlation
was found.

\section{Color Reconnection}

\begin{figure}
\begin{center}
\epsfig{file= 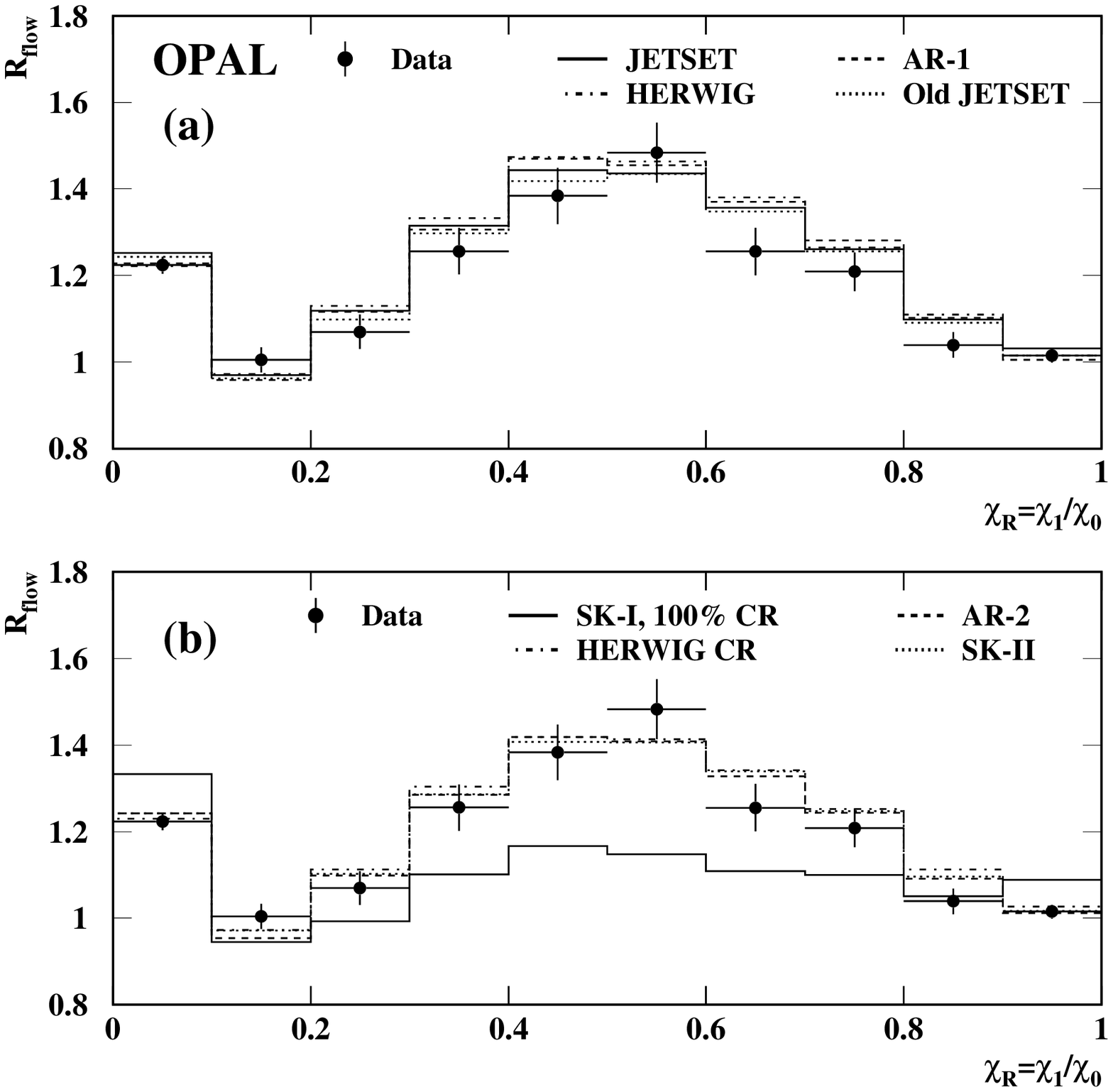,width=7.7cm} 
\epsfig{file= 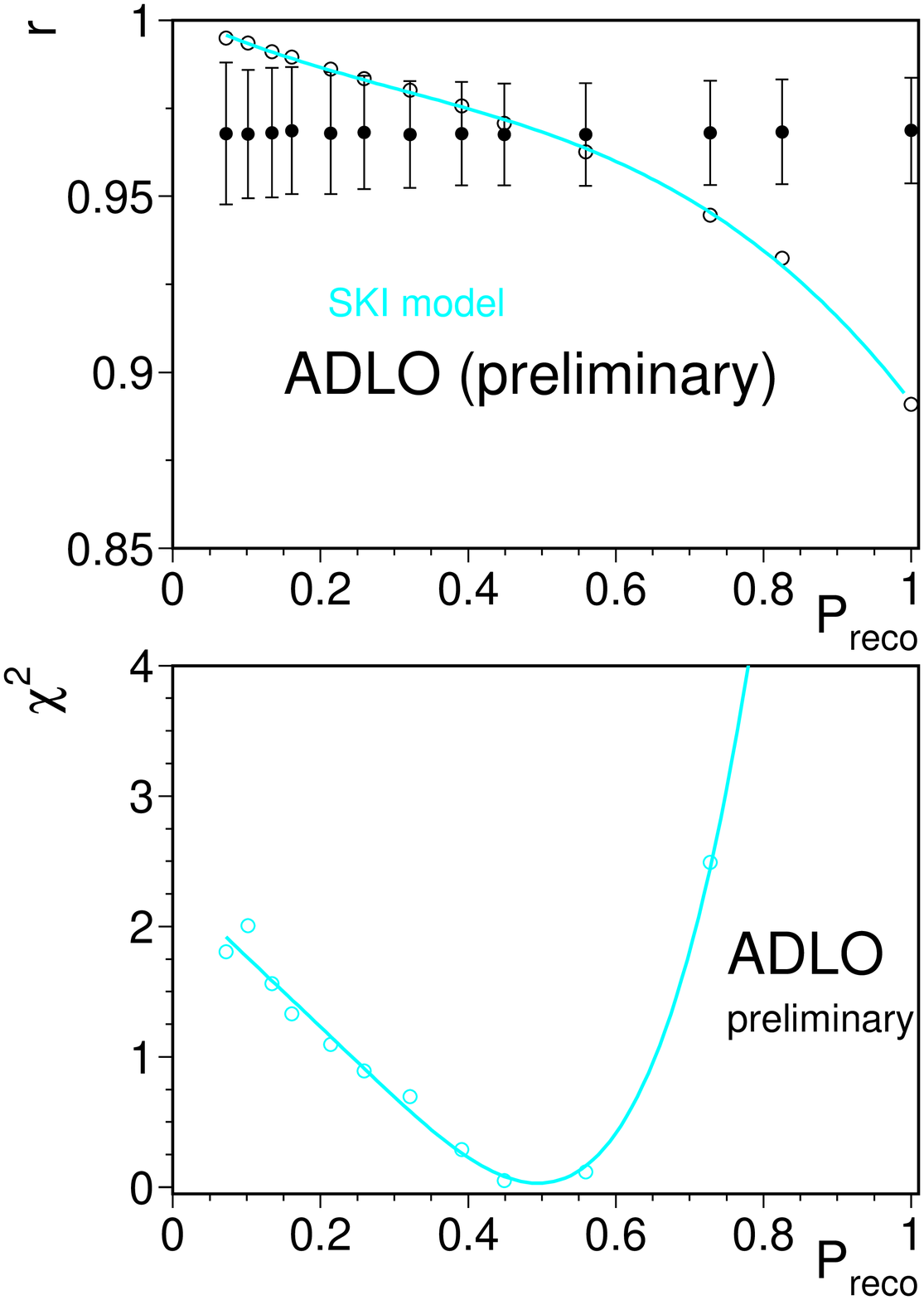,width=4.8cm} 
\end{center}
\caption{Left: Comparison of the ratio of particle flow in
the { \it inter} to the {\it intra} W regions between the data
and different Monte Carlo Models without (top) and with (bottom) color
reconnection. 
Right: Constraint on the strength of color reconnection from
the LEP combination
\label{fig-2}}
\end{figure}

The decay products of the two W bosons in hadronic W pair decays have a
significant space-time overlap as the separation of their decay
vertices at LEP2 energies is small compared to characteristic
hadronisation distance scales. It is therefore feasible that colored objects
from both W bosons interact during the hadronisation.
With the current understanding of non-perturbative QCD, such interference
can be estimated only in the context of specific models
\cite{bib:SK,bib:ARIADNE,bib:HERWIG}.
In general this models predict a changed charged multiplicity and a changed 
particle flow in the regions between jets from different W bosons ({\it inter}-W) compared to the 
regions between jets from the same W-boson ({\it intra}-W). The comparison of charged multiplicity
in hadronic W pair decays to that in semiletonic W pair decays is unfortunately not
precise enough to constrain the color reconnection models significantly.
The definition of regions between different jets in 4-jet events is non-trivial
and slightly different approaches have been used by the LEP experiments.
Figure~\ref{fig-2} (left) shows a comparison of the ratio of particle flow in
the {\it inter}- to the {\it intra}-W regions between the data and different Monte Carlo Models 
without and with color reconnection\cite{bib:opal}. In the SK-I model\cite{bib:SK} the strength of the
color reconnection can be adjusted. One can clearly see that the extreme 
scenario of 100\% CR (which is used as a benchmark and is not considered as a
realistic scenario by the authors) is excluded. The measurement can be
used to determine the most likely strength of color reconnection in the 
SK I model. The result of the LEP combination is shown in Figure~\ref{fig-2} on
the right\cite{lep-result}.

\end{document}